\documentclass[final,1p,number,sort&compress]{elsarticle}
\usepackage{amssymb, amsmath}
\usepackage[usenames, dvipsnames]{color}
\usepackage{esint}
\usepackage{epstopdf}
\usepackage{epsfig}
\newcommand{\be}{\begin{equation}}
\newcommand{\ee}{\end{equation}}
\newcommand{\ber}{\begin{eqnarray}}
\newcommand{\eer}{\end{eqnarray}}
\newcommand{\de}{\end{equation*}}
\newcommand{\cer}{\begin{eqnarray*}}
\newcommand{\der}{\end{eqnarray*}}

\newcommand{\diracslash}[1]{#1\llap{/\kern2pt}}
\begin{document}

\begin{frontmatter}

\title{Comparison of lower order and higher order nonclassicality in photon added and photon subtracted squeezed coherent states}

\author{Kishore Thapliyal$^\dagger$, Nigam Lahiri Samantray$^{\ddagger,}$\footnote{Presently at: INRIM, Strada delle Cacce 91, I-10135 Torino, Italy}, J. Banerji$^\ddagger$, Anirban Pathak$^{\dagger,}$\footnote{email: anirban.pathak@jiit.ac.in}}

\address{$^\dagger$Jaypee Institute of Information Technology, A-10, Sector 62, Noida, UP 201307, India,\\
$^\ddagger$Physical Research Laboratory, Navrangpura, Ahmedabad 380 009, India}

\begin{abstract}
Nonclassical properties of photon added and photon subtracted squeezed coherent states have been compared with specific focus on the higher order nonclassicalities, such as higher order squeezing, higher order sub-Poissonian photon statistics, higher order antibunching. It is observed that both  photon added and photon subtracted squeezed coherent states are highly nonclassical as they satisfy criteria for all of the above mentioned nonclassicalities and a set of other criteria including  negativity of Wigner function,   Klyshko's criterion and Agarwal's $(A_{3})$ parameter. Further, the amount of nonclassicality present in these two types of states has been  compared quantitatively using a measure of nonclassicality known as  nonclassical volume. Variation in the amount of nonclassicality with the number of photon(s) added/subtracted is also investigated, and it is found that the addition  of photons makes the squeezed coherent state  more nonclassical than what is done by the subtraction of photons.
\end{abstract}


\end{frontmatter}



\section{Introduction}
A state that cannot be expressed as a mixture of coherent states is referred to as a nonclassical state, and usually negativity of Glauber-Sudarshan $P$-function defines a nonclassical state \cite{glaub,sudar}. However,  $P$-function cannot be measured directly in  any experiment. This fact led to the construction of several other operational criteria for the detection of nonclassicality, such as negativity of Wigner function \cite{wig}, zeros of $Q$-function \cite{Sch-book}, and a set of moment based criteria \cite{Adam-criterion}. These operational criteria are only sufficient for the detection of nonclassicality. Thus, any of these criteria would fail to detect the presence of  nonclassicality in some cases. Still these criteria are used for long to study the possibilities of generation of  nonclassical states and to propose their applications (see \cite{dodo2} and references therein). However, the interest on nonclassical states has been intensified in the recent past with the introduction of several new applications of nonclassical states \cite{LIGO1,LIGO2,CV-qkd-hillery,sc1,teleportation-of-coherent-state,antibunching-sps,Ekert-protocol,Bennet1993,densecoding}.
For example, applications of squeezed state are reported in the detection of gravitational waves in the famous LIGO experiment \cite{LIGO1, LIGO2} and in the realization of schemes for continuous variable quantum key distribution \cite{CV-qkd-hillery} in particular, and schemes for continuous variable quantum communication in general \cite{sc1},
teleportation of coherent states \cite{teleportation-of-coherent-state},
etc., antibunching is reported to be useful in characterizing single photon
sources \cite{antibunching-sps}, entangled states are found to be essential for the realization of  protocols
for discrete \cite{Ekert-protocol} and continuous variable quantum
cryptography \cite{CV-qkd-hillery}, quantum teleportation \cite{Bennet1993},
dense-coding \cite{densecoding}, etc. As a consequence of these useful developments, in the recent years, nonclassical states of light have been recognized as a valuable resource in quantum information processing  and quantum optics. Many experimental schemes have been proposed to generate nonclassical states. Two basic operations that can lead to nonclassical states are adding photons to or subtracting photons from a quantum state \cite{sc8}. In what follows, we aim to investigate the effect of these two processes on the squeezed coherent states.

Today, there exists a rich variety of
nonclassical states of light, either in reality or (at least) in
principle \cite{dodo2}. Our objective is to search for  nonclassicality in these states or superpositions of these states so that the states can be used for the applications mentioned above. Specifically, we would like to study the nonclassical nature of light present in photon added/subtracted squeezed coherent states through some particular manifestations of the nonclassical nature of light, such as lower order and higher order photon antibunching, sub-Poissonian distribution of photon numbers, squeezing and negativity of the Wigner function, Klyshko's criteria \cite{klys} and Agarwal's $A_{3}$ parameter \cite{Agarwal-Tara}.

In the studies on nonclassical states of light, operator ordering is considered to be one of the fundamental tasks as it plays a crucial role in obtaining expressions for various states of light and also in calculating the expectation values of the operators with respect to  these states. A simple, unified approach for arranging quantum operators into ordered products (normal ordering, antinormal ordering, Weyl ordering)  is the technique of {\it integration within an ordered product} (IWOP) \cite{sc11}. This technique is used here to perform a comparison of nonclassicality between photon-added squeezed coherent state (PASCS) and photon-subtracted squeezed coherent state (PSSCS). In the similar line, in \cite{jun}, a comparison of nonclassicality between photon-added and photon-subtracted squeezed vacuum states was performed, but the investigation reported there was restricted to lower order nonclassical phenomena. In Ref. \cite{Ashoka} nonclassical properties of photon subtracted squeezed state was also studied. The present  study aims to conduct a similar but much more rigorous comparison between PASCS and PSSCS with primary focus on higher order nonclassical phenomena. Specifically, in what follows, we  investigate the possibilities of observing various  lower order and higher order nonclassical properties of PASCS and PSSCS. The calculation is simplified considerably by using the IWOP technique. Earlier, some efforts had been made to study the lower order nonclassical properties of  PSSCS \cite{Wang-Fan12}. However, to the best of our knowledge, neither the existence of higher
order nonclassicality in PSSCS and PASCS, nor the variation of amount of nonlcassicality present in them and reflected through a quantitative measure of nonclassicality have yet been investigated. A set of recent studies on higher order nonclassicalities \cite{generalized-higher-order,HOAwithMartin,Maria-PRA-1,Maria-2,higher-order-PRL,with-Martin-hammar, symmetry-sanjib} have strongly established their relevance, especially in detecting  weak nonclassicalities. The effect of addition and/or subtraction of photon(s)  on the higher order nonclassicalities have been studied earlier in the context of photon added coherent state \cite{amit1, amit2, amit3} and a non-Gaussian entangled state generated by adding photons to both modes of a two-mode squeezed coherent state \cite{Wang-higher}. Possibility of observing higher order nonclassical effects through the addition of photons to coherent or thermal state was investigated in \cite{Ho-photon-added}. A similar study on the effect of addition/subtraction of photon(s)  t on the lower order nonclassicality has been reported for two-mode photon-added displaced squeezed states \cite{Hoai-Duc16}, Gaussian entangled state \cite{Dowling-Agarwal15}, squeezed coherent state \cite{Wang-Fan12}, and two-mode squeezed vacuum state \cite{Bartley15}. 

In \cite{Bartley15, Dowling-Agarwal15}  and in most of the other studies mentioned here, photon addition, photon subtraction, and their coherent superposition are considered as primary non-Gaussian operations (at the single photon level) that form one of the most powerful tools for controlling nonclassicality in general, and enhancement of entanglement in particular. Here, we report the role of these two types of operations in controlling nonclassicality in PASCS and PSSCS with special attention towards higher order nonclassicality. Apart from the recently reported applications of the nonclassical states, the task is interesting because of the possibility that the states studied here can be experimentally realized. This is so because, in \cite{exp,exp1} a photon added coherent state (PACS) proposed by Agarwal and Tara  \cite{gsatara} was experimentally generated, and thus the basic non-Gaussian operation required for the generation of PASCS and PSSCS is already achieved. 

The remaining part of the paper is organized as follows. In Sec. \ref{sec:PASCS-PSSCS}, we introduce the quantum states of our interest (i.e., PASCS and PSSCS) and obtain analytic expressions for the required normalization constants. In Sec. \ref{sec:moments}, we report two closed form analytic expressions for $\langle a^{\dagger p} a^{q}\rangle$, where the expectation values are computed with respect to (a) PASCS and (b) PSSCS. In subsequent sections these expressions are used to investigate the existence of nonclassicality using various criteria that are based on moments of the annihilation and creation operators. Specifically, in Sec. \ref{hoaandhosps}, we use a set of moment-based criteria and the expressions obtained in Sec. \ref{sec:moments} to study the existence of higher order antibunching and higher order sub-Poissonian photon statistics in PASCS and PSSCS. Subsequently, in Sec. \ref{sec:pnandklyshko}, we report photon number distribution for PASCS and PSSCS and demonstrate that they satisfy Klyshko's criterion for nonclassicality, which requires very limited knowledge of photon number distribution. Nonclassical properties of PASCS and PSSCS reflected through Agarwal's criterion are reported in Sec. \ref{sec:Agarwal} and  the existence of Hong-Mandel type higher order squeezing in these states are reported in Sec. \ref{HOS}. In Sec, \ref{sec:Wigner}, we first report analytic expressions for Wigner function for PASCS and PSSCS and investigate the existence of nonclassicality in these states through the negative values of the Wigner function, then we provide a quantitative measure of nonclassicality present in these states using nonclassical volume and use the same for comparison of nonclassicality present in these states as the number of photons added/subtracted is varied. Finally, the paper is concluded in Sec. \ref{sec:conclusion}.

\section{Quantum states of our interest}\label{sec:PASCS-PSSCS}
This paper is focused on two types of quantum states, namely PASCS and PSSCS. Before, we investigate nonclassical properties of these states, it would be apt to properly define these states and to obtain the analytic expressions for the corresponding normalization constants. To begin with let us define PASCS.
\subsection{Definition of PASCS:}
The PASCS is defined as 
\be
\vert\psi\rangle=N_{+}\left(\alpha,z,m\right){\hat{a}^{\dagger m}} \vert\alpha,z\rangle,\label{eq:pascs}
\ee
where $\vert\alpha,z\rangle=D(\alpha)S(z)\vert 0\rangle$ is the squeezed coherent state (SCS), $m$ is the number of photons added, $\vert 0\rangle$ is the vacuum state, $D(\alpha)=\exp(\alpha a^{\dagger}-\alpha^{*} a)$ is the displacement operator, and $S(z)=\exp(\frac{1}{2}z {a^\dagger }^2 -\frac{1}{2}z^* a^2 )$ is the squeeze operator. Here, $ z =  r\exp(i\phi)$, where $r$ is the squeezing parameter and $\phi$ is the squeeze angle. The normalization constant $ N_{+}\left(\alpha,z,m\right)$ is determined from the normalization condition  $ \langle\psi\vert\psi\rangle = 1$. Using the IWOP technique, we get
\be
N_{+}\left(\alpha,z,m\right) = \left[\left(\frac{\sinh 2r}{4} \right)^{m} \sum_{l=0}^{m} \frac{\left(m!\right)^{2} \left(2\coth r\right)^{l}}{l! \{\left(m-l
\right)!\}^2} \left|H_{m-l}( A)\right|^2 \right]^{-1/2},\label{eq:pascs-nor}
\ee
where
\be
A=i\frac{\alpha\exp(-i\phi/2)}{\sqrt{\sinh 2 r }}.\nonumber
\ee
For $ z=0 $, $ N_{+}\left(\alpha,z,m\right) $ becomes the normalization constant for photon added coherent state \cite{gsatara}
\be
N_{+}\left(\alpha,0,m\right) = \left[m!L_m\left(-\left|\alpha\right|^2\right)\right]^{-1/2},\nonumber
\ee
and for $ \alpha=0 $, $N_{+}\left(\alpha,z,m\right)$  is the normalization constant for photon added squeezed vacuum state \cite{quesne}
\be
N_{+}\left(0,z,m\right) = \left[m!\left(\cosh r\right)^{m}P_{m}\left(\cosh r\right)\right]^{-1/2}.\nonumber
\ee

In a similar fashion, we may define PSSCS as follows.

\subsection{Definition of PSSCS:}
The PSSCS is defined as 
\be 
\vert\psi\rangle = N_{-}\left(\alpha,z,m\right)\hat{a}^{m} \vert\alpha,z\rangle, \label{eq:psscs}
\ee 
where $ N_{-}\left(\alpha,z,m\right) $ is the normalization constant,
\be
N_{-}\left(\alpha,z,m\right) = \left[\left(\frac{\sinh 2r}{4} \right)^{m} \sum_{l=0}^{m} \frac{\left(m!\right)^{2} \left(2\tanh r\right)^{l}}{l! \{\left(m-l
\right)!\}^2} \left|H_{m-l}( A)\right|^2 \right]^{-1/2}.\label{eq:psscs-nor}
\ee
For $ z=0 $ , $ N_{-}\left(\alpha,z,m\right)=\vert\alpha\vert^{-m}$ as expected, whereas for $\alpha=0$, $ N_{-}\left(\alpha,z,m\right)$ is the normalization constant for photon subtracted squeezed vacuum state \cite{quesne}
\be
N_{-}\left(0,z,m\right)=m!(-i\sinh r)^{m}P_{m}(i\sinh r).\nonumber
\ee
\section{Analytic expressions for expectation values}\label{sec:moments}

In Ref. \cite{Agarwal-Tara}, a moment-based criterion written in the matrix form was proposed for the detection of nonclassicality, which was further extended to an entanglement criterion \cite{vogel's-ent-criterion}. Later, it was realized that most of the well known criteria for nonclassicality can be expressed as functions of moments of annihilation and creation operators \cite{Adam-criterion,Adam-generalizes-vogel}. Thus, if one can obtain an analytic expression for $\langle a^{\dagger p} a^{q}\rangle$ for a particular state, one may use the same to study the possibilities of observing various types of nonclassicality through different moment-based criteria of nonclassicality (cf. Table I of \cite{Adam-criterion}).  Keeping this in mind, in this section, we first calculate $\langle a^{\dagger p} a^{q}\rangle$. In the following sections, we will use the expressions obtained in this section to compare nonclassical properties of PASCS and PSSCS.
\subsection{Expectation values with respect to PASCS:}
Using Eqs. (\ref{eq:pascs})-(\ref{eq:pascs-nor}) and a few steps of computation, we obtain 
\be
\langle a^{\dagger p} a^{q}\rangle =  N^{2}_{+}\left(\alpha,z,m\right) \langle\beta\vert S^{\dagger}(z) a^{m} {a^{\dagger}}^{p} a^{q} {a^{\dagger}}^{m} S(z) \vert\beta\rangle,\nonumber
\ee
i.e.,
\ber
\langle a^{\dagger p} a^{q}\rangle &=&  N^{2}_{+}\left(\alpha,z,m\right) \exp\left[i\phi (p-q)/2\right] \sum_{j=0}^{\min (p,q)} \frac{\left(-1 \right )^j p! q!(-i)^p i^q}{j!(p-j)!(q-j )!} \left(\frac{\sinh 2r}{4}\right)^{m-j+\frac{p+q}{2}}\nonumber \\ && \times  \sum_{l=0}^{m-j+\min (p,q)} \frac{(2\coth r )^l}{l!} \frac{(m+p-j )!(m+q-j )!}{(m+p-j-l)! (m+q-j-l )!} \nonumber \\ && \times H_{m+p-j-l}(A) H_{m+q-j-l}(A^* ).\label{eq:mom-pascs}
\eer

For $p=q$, it is found that

\ber
\langle a^{\dagger p} a^{p}\rangle &=&  N^{2}_{+}\left(\alpha,z,m\right)\sum_{j=0}^{p} \frac{\left(-1 \right )^j \left(p!\right)^{2}}{j!\left\{(p-j)!\right\}^{2}} N^{-2}_{+}\left(\alpha,z,m-j+p\right).\label{eq:mom-pascs-p}
\eer

\subsection{Expectation values with respect to  PSSCS:}
Similarly, using Eqs. (\ref{eq:psscs})-(\ref{eq:psscs-nor}) and a few steps of simplification, we obtain 
\ber
\langle a^{\dagger p} a^{q}\rangle &=& N^{2}_{-}\left(\alpha,z,m\right) \exp\left[-i\phi (p-q)/2\right] (-i)^p i^q\nonumber\\&& \times \sum_{l=0}^{m+\min (p,q)} \frac{(2\tanh r )^l}{l!} \frac{(m+p)!(m+q)!}{(m+p-l )! (m+q-l )!}\left(\frac{\sinh 2r}{4}\right)^{m+\frac{p+q}{2}} \nonumber \\ && \times H_{m+p-l}(A) H_{m+q-l}(A^* ).\label{eq:mom-psscs}
\eer

For $p=q$, it reduces to

\ber
\langle a^{\dagger p} a^{p}\rangle &=&  \frac{N^{2}_{-}\left(\alpha,z,m\right)}{N^{2}_{-}\left(\alpha,z,m+p\right)}.\label{eq:mom-psscs-p}
\eer

We have already mentioned that the analytic expressions for $\langle a^{p} a^{\dagger q} \rangle$ can be used to study the possibilities of existence and variation of different signatures of nonclassicality. In the following sections, this particular point is illustrated through some explicit examples. To begin with, in the next section we  discuss nonclassicality reflected through Mandel's $Q$ parameter, and criteria for higher order antibunching and higher order sub-Poissonian photon statistics. 

\section{Mandel's $Q$ parameter, higher order antibunching and higher order sub-Poissonian photon statistics}\label{hoaandhosps}
Mandel's $Q$ parameter is defined as

\be
Q=\frac{\langle a^{\dagger 2}a^{2}\rangle}{\langle a^{\dagger} a\rangle}-\langle a^{\dagger} a\rangle.\nonumber
\ee
Negative values of $Q$ parameter imply negative values for $P$ function and thus it provides us a witness for nonclassicality which can be computed easily using Eq. (\ref{eq:mom-pascs-p}) for PASCS and Eq. (\ref{eq:mom-psscs-p}) for PSSACS. However, we do not discuss it separately, as it can be obtained as a special case of higher order nonclassical criteria. Specifically, a variant of Lee's criterion for higher order antibunching \cite{CTLee} was  proposed by Pathak and Garcia \cite{HOAwithMartin}  as follows: A state will be considered as  $(n-1)$th order antibunched if it satisfies  
\begin{equation}
\begin{array}{lcl}
D(n-1)=\left\langle a^{\dagger n}a^{n}\right\rangle -\left\langle a^{\dagger}a\right\rangle ^{n} & < & 0.\end{array}\label{hoa}\end{equation}
The negative values of $D(1)$ ensure negativity of $Q$ parameter or equivalently conventional antibunching. Thus, if the above criterion is satisfied for $n=2$ that would mean antibunching, whereas the satisfaction of the criterion for $n\geq3$ would imply higher order antibunching of $(n-1)$th order.  As we are interested in both lower and higher order antibunching, in this section, we prefer to use Eq. (\ref{hoa}) for our investigation. It is also noteworthy here that using Eqs. (\ref{eq:mom-pascs-p}) and (\ref{eq:mom-psscs-p}) one can also calculate a higher order counterpart of $Q$ parameter reported in \cite{Qm}.

\begin{figure}
\centering{}\includegraphics[scale=0.6]{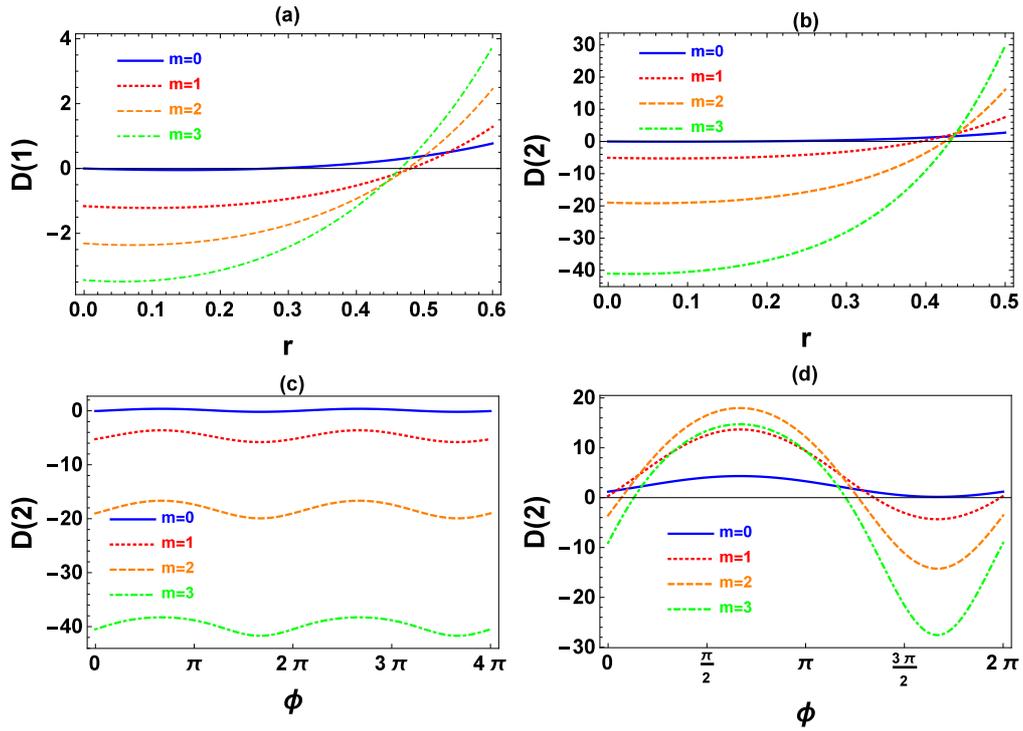}
\protect\caption{\label{fig:HOA-PASC} (Color online) Lower and higher-order antibunching is shown to vary with squeezing parameters $r$ and $\phi$ for PASCS in (a)-(b) and (c)-(d), respectively. The coherent state parameter is $\alpha=\sqrt{\frac{2}{3}}\exp\left(\frac{i \pi}{3}\right)$ in all the plots. In (a) and (b), $\phi=0$; while $r=0.1$ and 0.4 in (c) and (d), respectively. In (c), it can be observed that the variation of nonclassicality observed is periodic in nature with periodicity $2\pi$.  This periodicity is observed in general, and in (d) and all the subsequent figures, where we have shown dependence of a nonclassicality witness on the phase parameter $\phi$, we have restricted $\phi$ in the range $\phi=0$ to $2\pi$.}
\end{figure}

\begin{figure}
\centering{}\includegraphics[scale=0.6]{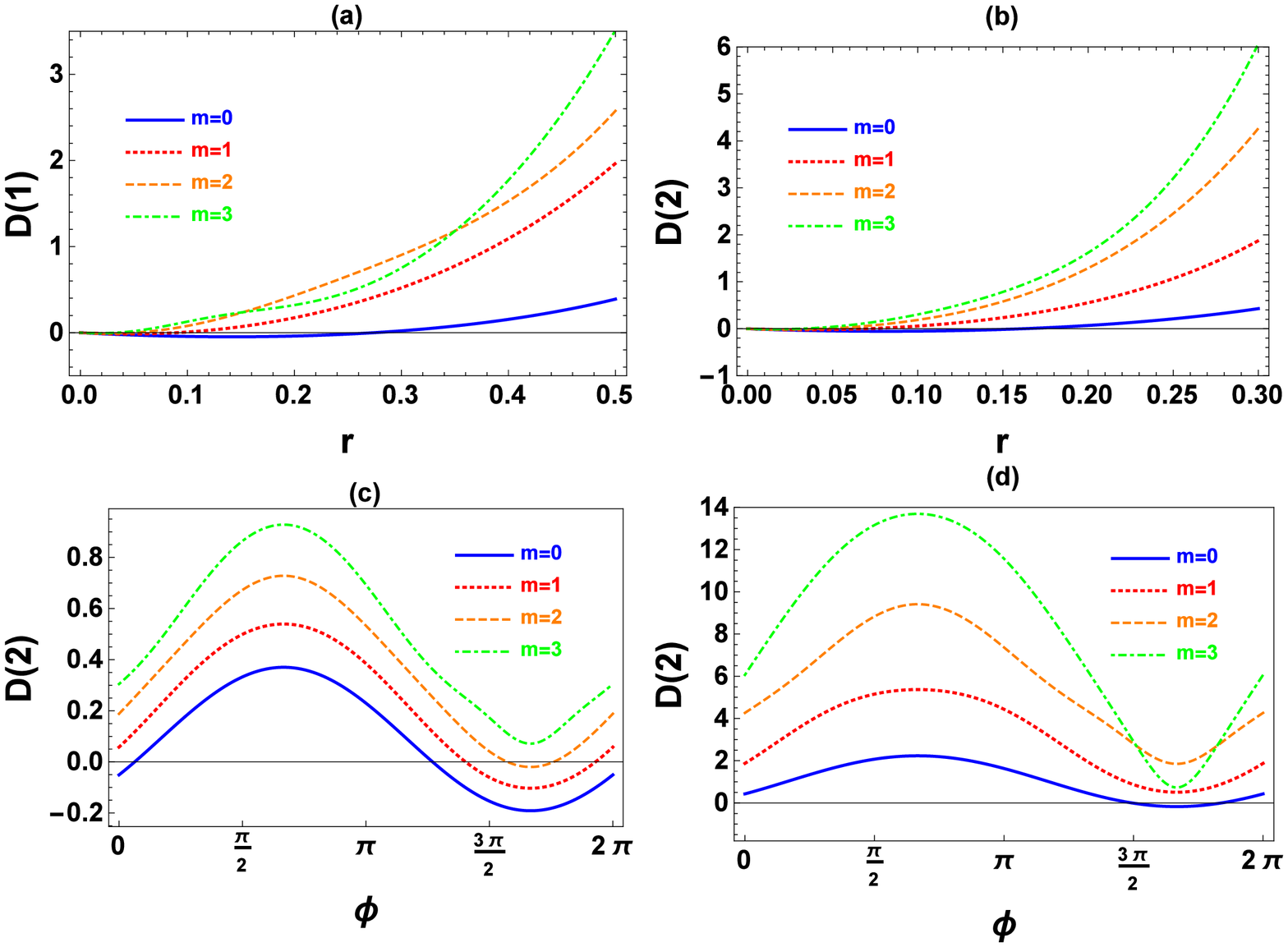}
\protect\caption{\label{fig:HOA-PSSC} (Color online) Lower and higher-order antibunching is shown to vary with squeezing parameters $r$ and $\phi$ for PSSCS in (a)-(b) and (c)-(d), respectively. The coherent state parameter is $\alpha=\sqrt{\frac{2}{3}}\exp\left(\frac{i \pi}{3}\right)$ in all the plots. In (a) and (b), $\phi=0$; while $r=0.1$ and 0.4 in (c) and (d).}
\end{figure}

Using Eq. (\ref{eq:mom-pascs-p}) in Eq. (\ref{hoa}) for $n=2$ and 3, we have obtained variation shown in Figure \ref{fig:HOA-PASC}. It can be noted that for $m=0$ the results reduces to SCS, which show both lower and higher order antibunching only for small values of squeezing parameter $r$ for specific values of $\phi$. With increase in the number of added photon(s), depth of nonclassicality is observed to increase, which is also observed to depend on the phase  of the squeezing parameter. It is also important to note that all the witnesses of nonclassicality studied here are found to display  periodic behavior for phase angle $\phi$ with period $2\pi$ as illustrated in Figure \ref{fig:HOA-PASC} (c) (in the remaining figures we have not plotted nonclassicality witnesses for $\phi=2\pi-4\pi$ region as that would be just repetition of the plot shown for $\phi=\pi-2\pi$ region). Similarly, in Ref. \cite{Wang-Fan12}, the authors observed periodicity of $\pi$ in case of photon number distribution.
A similar investigation  is performed for the PSSCS by substituting Eq. (\ref{eq:mom-psscs-p}) in Eq. (\ref{hoa}), and both lower order and higher antibunching are observed. Lower order ($n=2$) and higher order ($n=3$) antibunching are shown in Figure \ref{fig:HOA-PSSC}. Here, the nonclassicality present in the SCS can be easily observed. Also, the subtraction of photon(s) is found to reduce the nonclassicality present in the SCS.

Using the moments calculated in Eqs. (\ref{eq:mom-pascs}) and (\ref{eq:mom-psscs}) the photon statistics of PASCS and PSSCS can also be checked and compared with Poissonian photon statistics. This is expected to reveal nonclassical characters as both lower order and higher order sub-Poissonian photon statistics are known to be witnesses of nonclassicality in the sense that their presence ensures negativity of $P$ function, but the converse is not true.   The higher order criterion for nonclassical behavior of photon statistics, i.e., criterion for higher order sub-Poissonian photon statistics, is given as  
\begin{equation}
\begin{array}{lcll}
d(n-1)&=&\sum_{r=0}^{n}\sum_{k=1}^{r}S_{2}(r,k)\,^{n}C_{r}(-1)^{r}\langle a^{\dagger k}a^{k}\rangle\langle a^{\dagger}a\rangle^{n-r}&\\
&-&\sum_{r=0}^{n}\sum_{k=1}^{r}S_{2}(r,k)\,^{n}C_{r}(-1)^{r}\langle a^{\dagger}a\rangle^{k+n-r}
&<0,\end{array}\label{eq:cond-HOSPS}\end{equation}
where $S_{2}(r,k)$ is used for the Stirling number of the second kind.

Before we proceed with the discussion of the higher order sub-Poissonian photon statistics of PASCS and PSSCS, it is worth commenting that antibunching and sub-Poissonian photon statistics are shown to be two independent nonclassical behaviors in the recent past (cf. Table I in Ref. \cite{amit2}). The presence of one does not ensure the presence of the other. Due to this reason we study higher order sub-Poissonian photon statistics separately here. Further, it is also established in the past that higher order antibunching and sub-Poissonian photon statistics can exist irrespective of whether the lower order of the corresponding phenomena exists or not \cite{amit2,rita}.

\begin{figure}
\centering{}\includegraphics[scale=0.6]{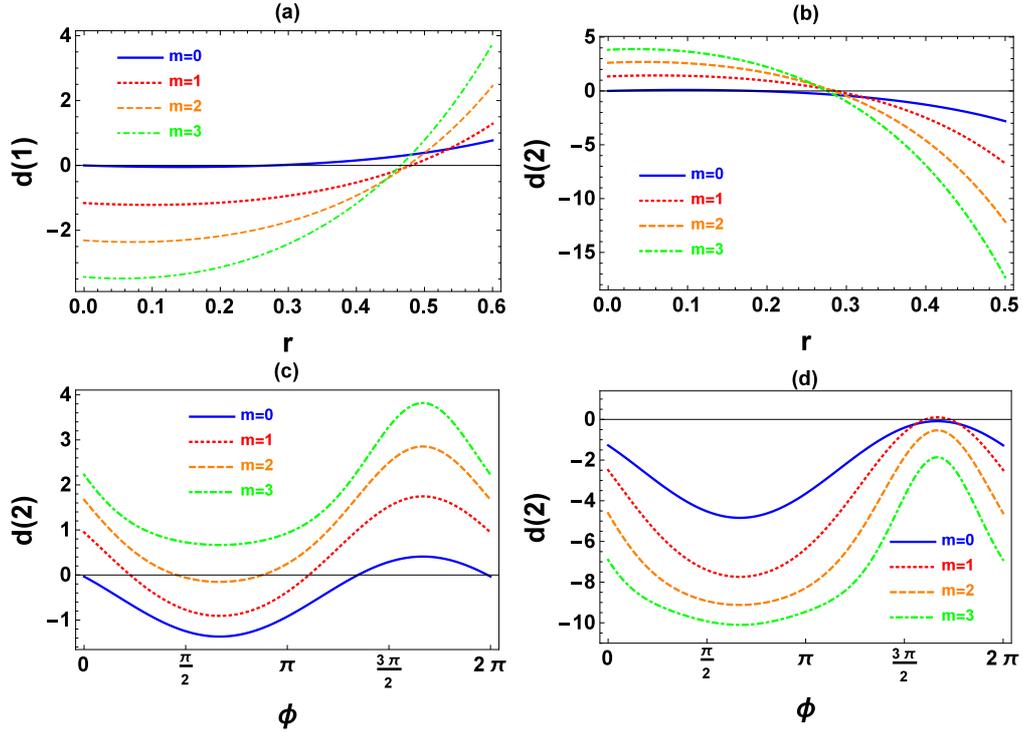}
\protect\caption{\label{fig:HOSPS-PASC} (Color online) Lower and higher order sub-Poissonian photon statistics with different values of photon addition (i.e., from 0-3) are shown for PASCS. (a) and (b) illustrate the lower and second order photon statistics characterized by $d\left(1\right)$ and $d\left(2\right)$, respectively.  For both these plots, squeezing parameter $\left(\phi\right)$ is chosen to be zero. In (c) and (d), the dependence of $d\left(2\right)$ on phase angle of squeezing parameter $\left(\phi\right)$ is established with squeezing parameter $r=0.2$ and 0.4, respectively. The coherent state parameter is $\alpha=\sqrt{\frac{2}{3}}\exp\left(\frac{i \pi}{3}\right)$ in all the plots.}
\end{figure}

Using Eq. (\ref{eq:mom-pascs-p}) in Eq. (\ref{eq:cond-HOSPS}), we have obtained sub-Poissonian photon statistics (i.e., for $n=2$) in PASCS, which is found to decay with increasing squeezing parameter (cf. Figure \ref{fig:HOSPS-PASC}). In the higher order sub-Poissonian photon statistics, a converse nature is observed, i.e., the presence of nonclassicality at the higher values of squeezing operator. This fact establishes the motivation to study higher order nonclassicality as \textit{it demonstrated nonclassicality in the region where lower order nonclassicality criteria failed to detect it}. It is also shown to depend upon the phase of the squeezing parameter.

\begin{figure}
\centering{}\includegraphics[scale=0.6]{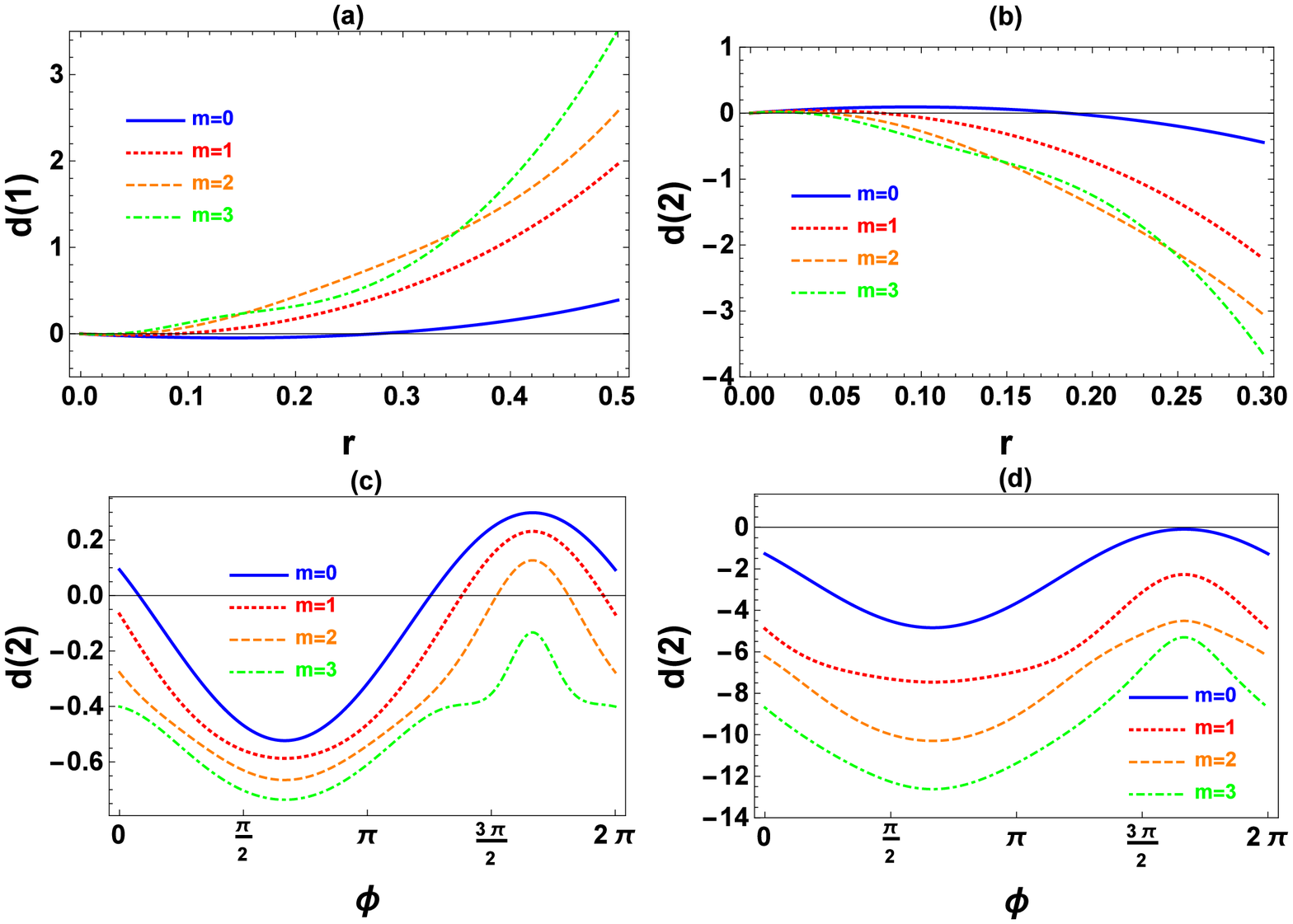}
\protect\caption{\label{fig:HOSPS-PSSC} (Color online) Lower and higher order sub-Poissonian photon statistics with different values of photon subtraction (i.e., from 0-3) are shown for PSSCS. (a) and (b) illustrate the lower and second order photon statistics characterized by $d\left(1\right)$ and $d\left(2\right)$, respectively. For both these plots, squeezing parameter $\left(\phi\right)$ is chosen to be zero. In (c) and (d), the dependence of $d\left(2\right)$ on phase angle of squeezing parameter $\left(\phi\right)$ is established with squeezing parameter $r=0.2$ and 0.4, respectively.  The coherent state parameter is $\alpha=\sqrt{\frac{2}{3}}\exp\left(\frac{i \pi}{3}\right)$ in all the plots.}
\end{figure}

The outcome of a similar study for PSSCS using Eq. (\ref{eq:mom-psscs-p}) in Eq. (\ref{eq:cond-HOSPS}) is illustrated in Figure \ref{fig:HOSPS-PSSC}. Apart from the facts observed for PASCS, higher order nonclassicality is observed to be more prominent in PSSCS when compared with PASCS.
Also, note that lower order antibunching and sub-Poissonian photon statistics show similar variation for both PASCS and PSSCS. However, higher orders of both these nonclassical behaviors have completely different dependence on the same set of parameters (cf. Figures \ref{fig:HOSPS-PASC} and \ref{fig:HOSPS-PSSC}).

\section{Photon number distribution $P_{n}$ and Klyshko's criteria}\label{sec:pnandklyshko}
\subsection{PASCS}
Photon number distribution  is described by the distribution of the probability $P_{n}$ of finding $n$ photons in the state $\vert\psi\rangle$. Thus, for PASCS, it can be  defined as

\ber
P_{n}&=&\vert\langle n\vert\psi\rangle\vert^2 \nonumber \\  &=& N^{2}_{+}\left(\alpha,z,m\right) \vert\langle n\vert {a^{\dagger}}^{m}\vert\alpha,z\rangle\vert^2.
\eer
Clearly, for $n<m$, $P_{n}=0$. If  $P_{n}=0$ for any $n$, it can be viewed as a hole in the photon number distribution and the same is known to be a manifestation of nonclassical character \cite{holeburn}. This is so because Glauber-Sudarshan $P$-function in such a case cannot be interpreted as a classical probability distribution \cite{holeburn}. In other words, the non-Gaussian operation of photon addition causes hole burning. For $n\geq m$, we have the expression
\ber
P_{n}& = & \frac{N^{2}_{+}\left(\alpha,z,m\right) n!}{\{\left(n-m\right)!\}^{2} \cosh r} \exp\left\{-\vert\beta\vert^{2} -\Re\left(e^{-i\phi}\beta^{2}\right)\tanh r\}\right\} \nonumber \\ && \times \left(\frac{\tanh r}{2}\right)^{n-m}  \left|H_{n-m}\left(\frac{-i\beta e^{-i\phi/2}}{\sqrt{\sinh 2r}}\right)\right|^{2},\label{eq:Pn-PASCS}
\eer
where $\beta= \alpha\cosh r -\alpha^{*} e^{i\phi} \sinh r$.

\subsection{PSSCS:}
\begin{figure}
\centering{}\includegraphics[scale=0.6]{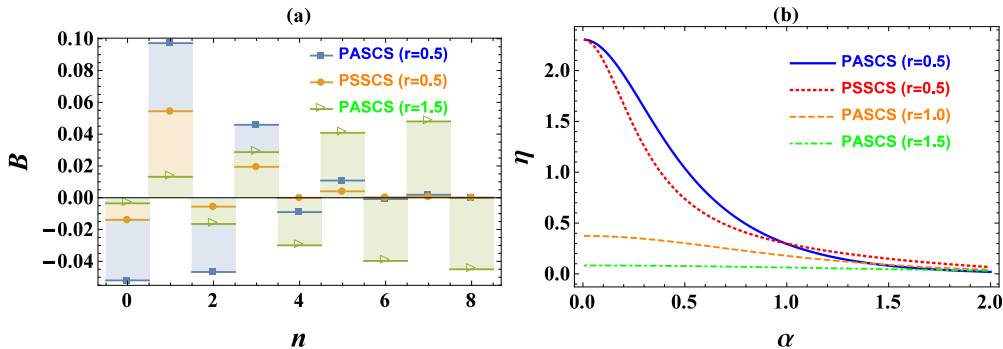}
\protect\caption{\label{fig:eta-Klysko} (Color online) (a) Nonclassiality of both PASCS and PSSCS using Klyshko's criteria for different values of $r$ with $\alpha=1.0$. In all the cases where only variation corresponding to PASCS is shown, PSSCS shown the same variation. (b) Variation of $\eta$ parameter, which characterizes the nature of approximate single photon source, with coherent state parameter for different values of squeezing parameter $r$. In both (a) and (b), squeezing parameter $\phi=0$, and only single photon addition or subtraction is considered.}
\end{figure}

Similarly, for PSSCS,  we find
\ber
P_{n}&=& \frac{N^{2}_{-}\left(\alpha,z,m\right)}{n! \cosh r}  \exp\left\{-\vert\beta\vert^{2} -\Re\left(e^{-i\phi}\beta^{2}\right)\tanh r\}\right\} \nonumber \\ && \times \left(\frac{\tanh r}{2}\right)^{n-m}  \left|H_{n-m}\left(\frac{-i\beta e^{-i\phi/2}}{\sqrt{\sinh 2r}}\right)\right|^{2}.\label{eq:Pn-PSSCS}
\eer
The photon number distributions obtained so far  can now be used to investigate the existence of nonclassicality using Klyshko's criteria \cite{klys} which is given as
\begin{equation}
B=(n+2)P_{n}P_{n+2}-(n+1) P_{n}^2<0.\label{eq:klys}\end{equation}
This is an interesting criterion because to witness the nonclassicality it does not require information about the entire photon number distribution. It just requires the value of $P(n)$ for 3 consecutive values of $n$. Thus it reveals nonclassical characters of the state using only partial information about the state. Now, we may note that using $P_{n}$ from Eqs. (\ref{eq:Pn-PASCS}) and (\ref{eq:Pn-PSSCS}), one can easily obtain the  values of $B$, negativity of which would indicate the presence of  nonclassicality  in the quantum state. For specific values of parameters, we have shown variation of $B(n)$ in Figure \ref{fig:eta-Klysko} (a), which demonstrate the nonclassicality of both PASCS and PSSCS.

Further, it is well known that  a suitable candidate for the approximate single photon source should satisfy the criterion of antibunching and single photon sources are very important for quantum cryptography. We have already discussed the presence of antibunching in PASCS and PSSCS. Therefore, we may now introduce a quantitative measure $\eta$ to characterize the quality of the single photon source that can be prepared using these states. This parameter $\eta$ is the ratio of the number of single photon pulses to that with the number of photons more than one, i.e.,
\begin{equation}
\eta=\frac{P_{1}}{1-(P_{0}+P_{1})}.\label{eq:eta}\end{equation}
Using $P_{n}$ from Eqs. (\ref{eq:Pn-PASCS}) and (\ref{eq:Pn-PSSCS}) for characterizing the nature of the single photon source, we have demonstrated in Figure \ref{fig:eta-Klysko} (b). For higher values of the squeezing parameter both PASCS and PSSCS illustrate the same nature.

\section{Agarwal's criteria for nonclassicality}\label{sec:Agarwal}
A witness of nonclassicality $A_3$ using normally ordered moments written in matrix form \cite{Agarwal-Tara} is expressed as 
\begin{equation}
A_3=\frac{\det\, m^{(3)}}{\det\, \mu^{(3)}-\det\, m^{(3)}}<0,\label{eq:A3}\end{equation}
where \begin{equation}
m^{(3)}=\left[\begin{array}{ccc}
1&m_{1}&m_{2}\\
m_{1}&m_{2}&m_{3}\\
m_{2}&m_{3}&m_{4}\\\end{array}\right]\end{equation} and \begin{equation}
\mu^{(3)}=\left[\begin{array}{ccc}
1&\mu_{1}&\mu_{2}\\
\mu_{1}&\mu_{2}&\mu_{3}\\
\mu_{2}&\mu_{3}&\mu_{4}\\\end{array}\right].\end{equation}
Here, $m_{n}=\langle a^{\dagger n}a^{n}\rangle$ and $\mu_{n}=\langle \left(a^{\dagger}a\right)^{n}\rangle$. The $A_3$ parameter has a negative value for nonclassical states and becomes -1 for Fock states.

\begin{figure}
\centering{}\includegraphics[scale=0.6]{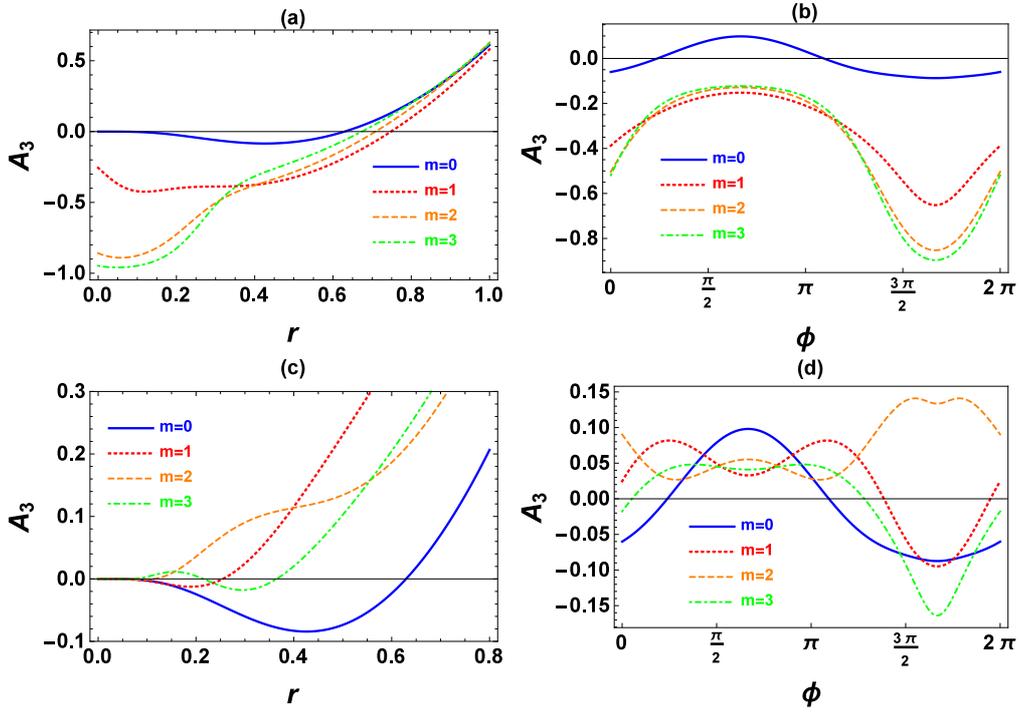}
\protect\caption{\label{fig:A3} (Color online) Variation of the nonclassicality witnessed using $A_{3}$ parameter with squeezing parameters $r$ and $\phi$ for PASCS and PSSCS in (a)-(b) and (c)-(d), respectively. The coherent state parameter is $\alpha=\sqrt{\frac{2}{3}}\exp\left(\frac{i \pi}{3}\right)$ in all the plots. In (a) and (c), $\phi=0$; while $r=0.3$ in (b) and (d).}
\end{figure}

We have used here the $A_3$ parameter to quantify the nonclassicality present in PASCS and PSSCS using Eqs. (\ref{eq:mom-pascs-p})  and (\ref{eq:mom-psscs-p}) in Eq. (\ref{eq:A3}), respectively. Variation of this parameter is shown in Figure \ref{fig:A3}. In Figure \ref{fig:A3} (a), it can be observed that with the photon addition the state become highly nonclassical, while with photon subtraction nonclassicality reduces (cf. Figure \ref{fig:A3} (c)). Additionally, in both the cases, the $A_3$ parameter depends upon the amount of squeezing as well as the phase of the squeezing parameter.

\section{Higher order squeezing}\label{HOS}

Higher order squeezing is studied using two independent criteria- Hillery's amplitude powered squeezing criterion \cite{HIlery-amp-sq} and Hong-Mandel higher order squeezing criterion \cite{Hong-Mandel1,HOng-mandel2}. Specifically, the Hillery-type squeezing considers the reduction of variances of amplitude powered quadratures compared with their coherent state counterparts as higher order squeezing \cite{HIlery-amp-sq}. Hong-Mandel-type higher order squeezing \cite{Hong-Mandel1,HOng-mandel2} takes into account higher order moments of usual quadrature $X=\frac{1}{\sqrt{2}}(a+a^{\dagger})$. Here, we restrict our discussion to the later type of higher order squeezing, which is known to provide the signature of nonclassicality only for even-orders. 
Specifically, it can be computed using the moments of annihilation and creation operators as signature of higher order squeezing can be found when following inequality involving higher order moment is satisfied \cite{amit2}

\begin{equation}
\begin{array}{l} 
\langle(\Delta X)^{n}\rangle=\sum_{r=0}^{n}\sum_{i=0}^{\frac{r}{2}}\sum_{k=0}^{r-2i}(-1)^{r}\frac{1}{2^{\frac{n}{2}}}t_{2i}\,^{r-2i}C_{k}\,^{n}C_{r}\,^{r}C_{2i}\langle a^{\dagger}+a\rangle^{n-r}\langle a^{\dagger k}a^{r-2i-k}\rangle<\left(\frac{1}{2}\right)_{\frac{n}{2}},\end{array}\label{eq:HOS1}\end{equation}
where $\left(x\right)_r$ is conventional Pochhammer symbol. This can be rearranged to introduce a simple witness of higher order squeezing as negative values of $S(n)$, where 
\begin{equation}
\begin{array}{lclcl}
S(n)&=&\frac{\langle(\Delta X)^{n}\rangle}{\left(\frac{1}{2}\right)_{\frac{n}{2}}}-1.\end{array}\label{eq:HOS2}\end{equation}

\begin{figure}
\centering{}\includegraphics[scale=0.6]{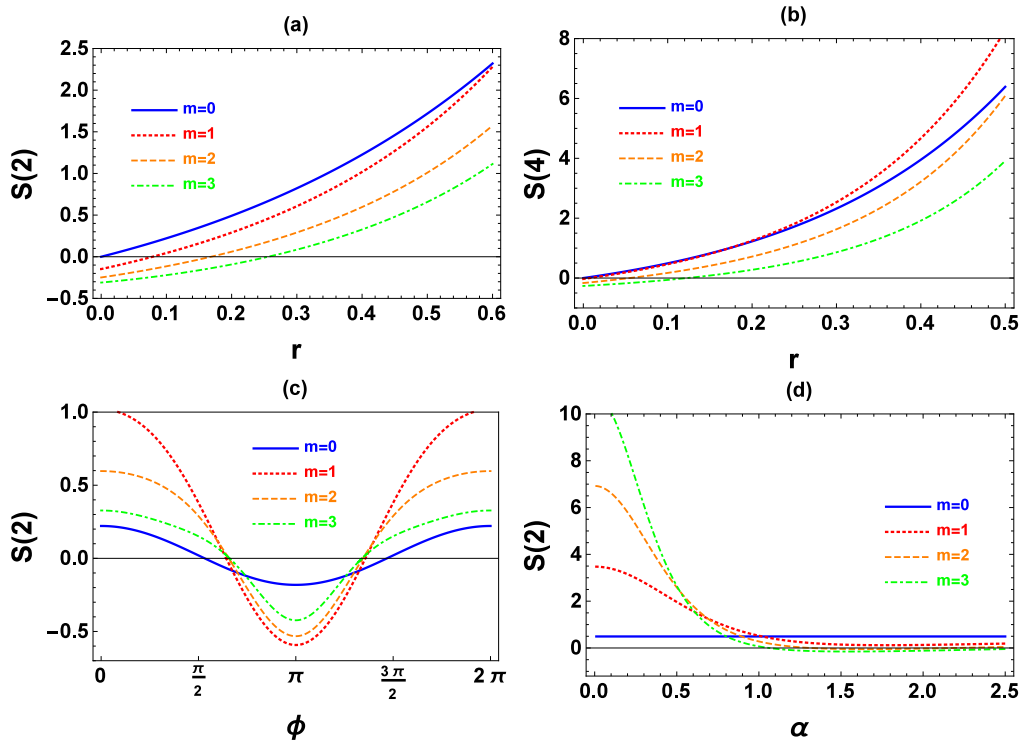}
\protect\caption{\label{fig:HOS-PASC} (Color online) Variation of the higher-order squeezing depicted in PASCS using Hong-Mandel criteria with squeezing parameters $r$ is shown in (a) and (b). In (c), the dependence on $\phi$ is established. (d) shows the variation with $\alpha$. In (a) and (c), $\alpha=1.2$, while $\alpha=1.4$ in (b). Also, $\phi=0$ in (a), (b) and (d), whereas $r=0.1$ in (c).}
\end{figure}

\begin{figure}
\centering{}\includegraphics[scale=0.6]{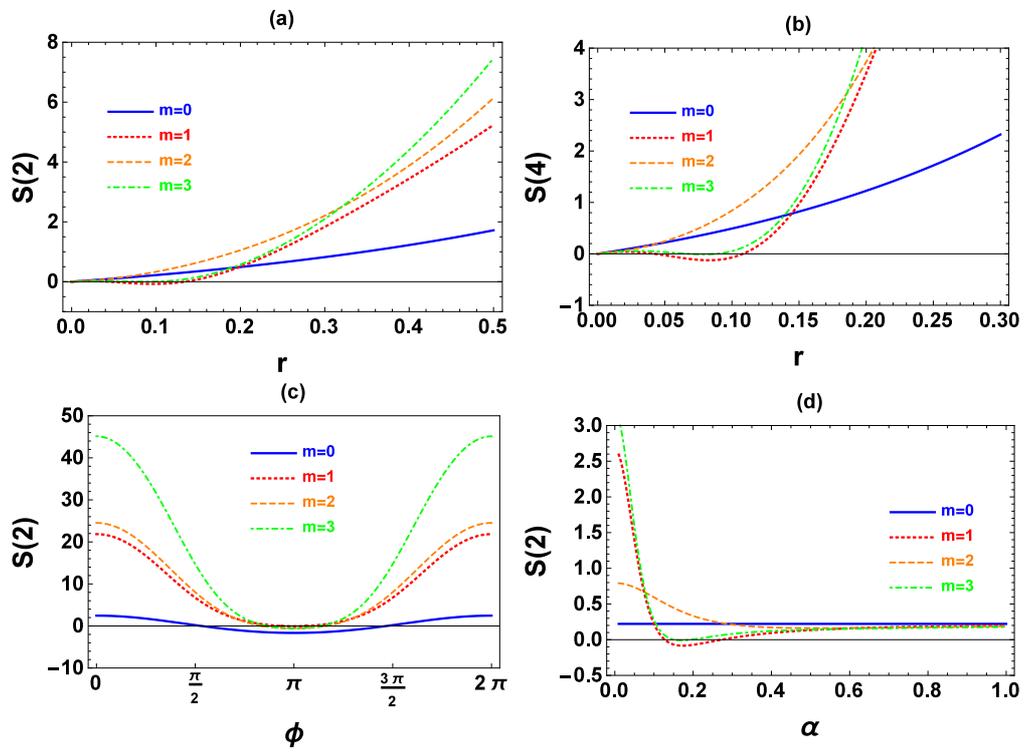}
\protect\caption{\label{fig:HOS-PSSC} (Color online) Higher-order squeezing depicted in PSSCS using Hong-Mandel criteria with squeezing parameters $r$ is shown in (a) and (b). In (c), the dependence on $\phi$ is established. (d) shows the variation with $\alpha$. In (a)-(c), $\alpha=0.2$. Also, $\phi=0$ in (a), (b) and (d), whereas $r=0.1$ in (c).}
\end{figure}

Substituting Eqs. (\ref{eq:mom-pascs})  and (\ref{eq:mom-psscs}) in Eq. (\ref{eq:HOS2}), variation of higher order squeezing in PASCS and PSSCS is obtained, and they are shown in Figures \ref{fig:HOS-PASC} and \ref{fig:HOS-PSSC}, respectively. The presence of squeezing is obtained in PASCS, and the existence of squeezing and the depth of the same is found to depend on the squeezing parameters. Further, the depth is found to increase with the number of photons added. In the case of PSSCS, squeezing was only present for certain number of photon subtractions. Further, PASCS and PSSCS show completely different natures with the coherent state parameter as PASCS (PSSCS) shows squeezing at large (only for small) values of the parameter.

\section{Wigner function and nonclassical volume}\label{sec:Wigner}
\subsection{PASCS}
The coherent state representation of Wigner function is defined as
\be
W\left(\gamma,\gamma^* \right ) = \frac{2}{\pi^{2}} \exp(2\vert\gamma\vert^{2}) \int \langle-\lambda\vert\hat{\rho}\vert\lambda\rangle \exp\{-2\left(\lambda\gamma^* -\lambda^* \gamma\right)\} {\mathrm{d}^{2}\lambda},
\ee
where $ \hat{\rho} = \vert\psi\rangle \langle\psi\vert$. Evaluating the term $ \langle - \lambda\vert\hat{\rho}\vert\lambda\rangle $ and on performing the integration over $\lambda$, $W\left(\gamma,\gamma^* \right )$ can be written in a product form as
\be
W(\gamma,\gamma^{*})= W_{SCS}\left(\gamma,\gamma^* \right ) F_{+}\left(m,\gamma,\gamma^* \right ),\label{eq:W-PASC}
\ee
where
\be
W_{SCS}\left(\gamma,\gamma^* \right ) = \frac{2}{\pi}\exp\{-(2\vert\gamma\cosh r-\gamma^{*}e^{i\phi}\sinh r-\beta\vert^{2})\},\nonumber
\ee
is the Wigner function of SCS, and 
\be
F_{+}\left(m,\gamma,\gamma^* \right ) = N^{2}_{+}\left(\alpha,z,m\right)\left(\frac{\sinh 2r}{4} \right )^{m}\sum_{l=0}^{m}\frac{\left(-2\coth r\right)^{l}}{l!}\left\{\frac{m!}{\left(m-l \right )!}\right\}^{2}\left| H_{m-l}\left(\frac{-ib_{+}}{\sqrt{\sinh 2r}}\right)\right|^{2}.\nonumber
\ee
Here,
\ber
b_{+} &=& e^{-i\phi/2}\left(2\gamma\cosh^{2} r-\beta\cosh r \right )-e^{i\phi/2}\left(\gamma^*\sinh 2r -\beta^*\sinh r\right).\nonumber
\eer

\subsection{PSSCS}
Similarly, the Wigner function of PSSCS can also be written in factorized form as

\be
W\left(\gamma,\gamma^* \right ) = W_{SCS}\left(\gamma,\gamma^* \right ) F_{-}\left(m,\gamma,\gamma^* \right ),\label{eq:W-PSSC}
\ee
where
\ber
F_{-}\left(m,\gamma,\gamma^* \right ) &=& N^{2}_{-}\left(\alpha,z,m\right)\left(\frac{\sinh 2r}{4} \right)^{m}\sum_{l=0}^{m}\frac{\left(-2\tanh r \right )^{l}}{l!}\left\{\frac{m!}{\left(m-l \right )!}\right\}^{2} \nonumber \\ && \times\left|H_{m-l}\left(\frac{-ib_{-}}{\sqrt{\sinh 2r}} \right )\right|^{2},\nonumber
\eer
with
\ber
b_{-} &=& e^{-i\phi/2}(\beta\cosh r-2\gamma\sinh^{2} r)-e^{ i\phi/2}(\beta^{*}\sinh r-\gamma^{*}\sinh 2r).\nonumber
\eer

\begin{figure}
\centering{}\includegraphics[scale=1.0]{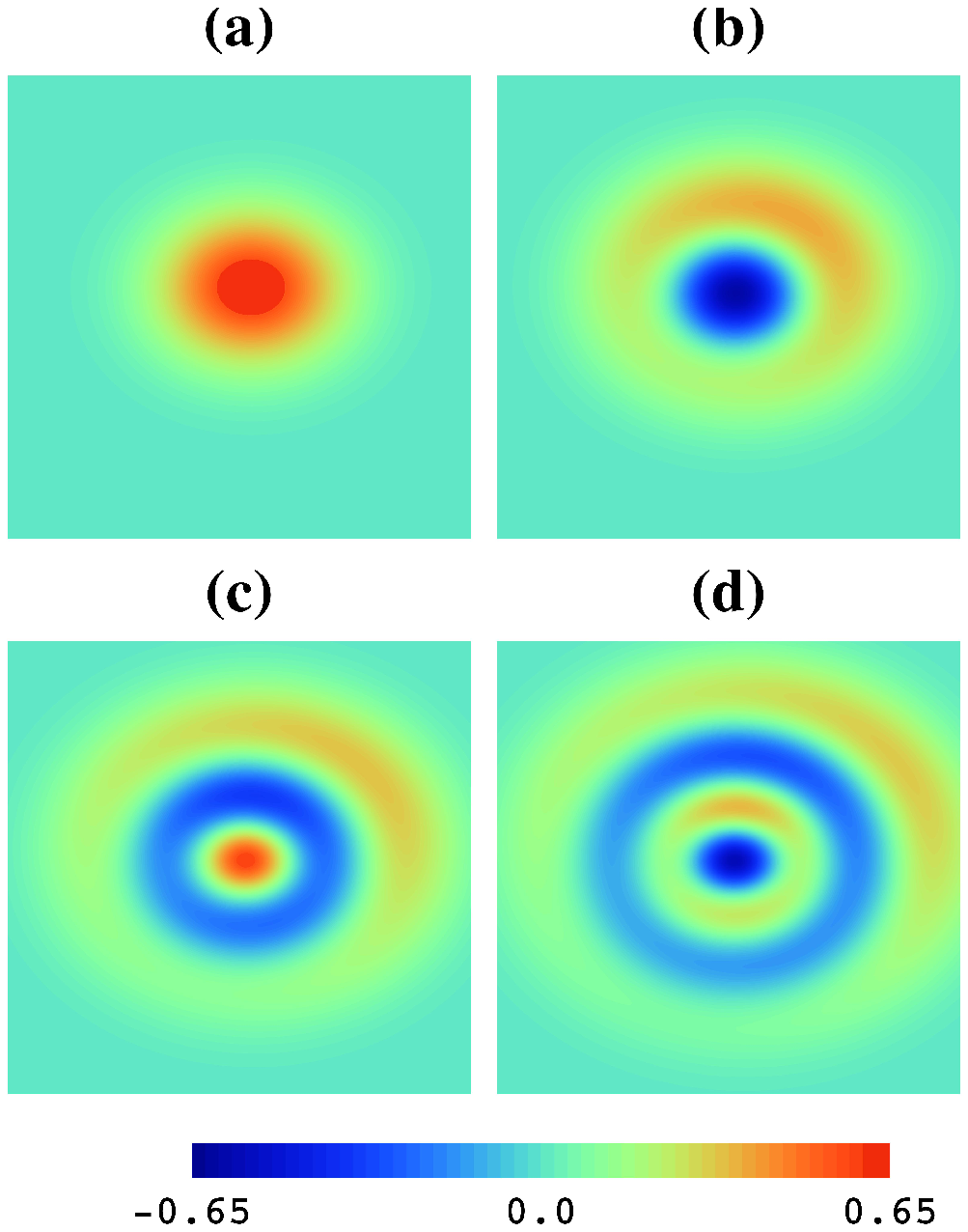}
\protect\caption{\label{fig:Wig-PASC} (Color online) Wigner function of PASCS for coherent state parameter $\alpha={\frac{1}{5}}\exp\left(\frac{i \pi}{3}\right)$ for photon addition of 0-3 in (a)-(d), respectively. Here, the squeezing parameters $r=0.1$ and $\phi=0$.}
\end{figure}

It is well known that negativity of the Wigner function  provides the signature of nonclassicality. Thus, negative regions present in a plot of a Wigner function would depict nonclassicality.  Keeping this in mind, variation of Wigner function of PASCS obtained in Eq. (\ref{eq:W-PASC}) is illustrated in Figure \ref{fig:Wig-PASC} with a repetitive photon addition from 0-3. For an even number of photon addition, the center is positive (cf. Figure \ref{fig:Wig-PASC} (a) and (c)), while it becomes negative for odd number of photon addition (cf. Figure \ref{fig:Wig-PASC} (b) and (d)). With increase in the number of photons added, the number of circular rings observed in the contour plots are found to increase.

\begin{figure}
\centering{}\includegraphics[scale=1.0]{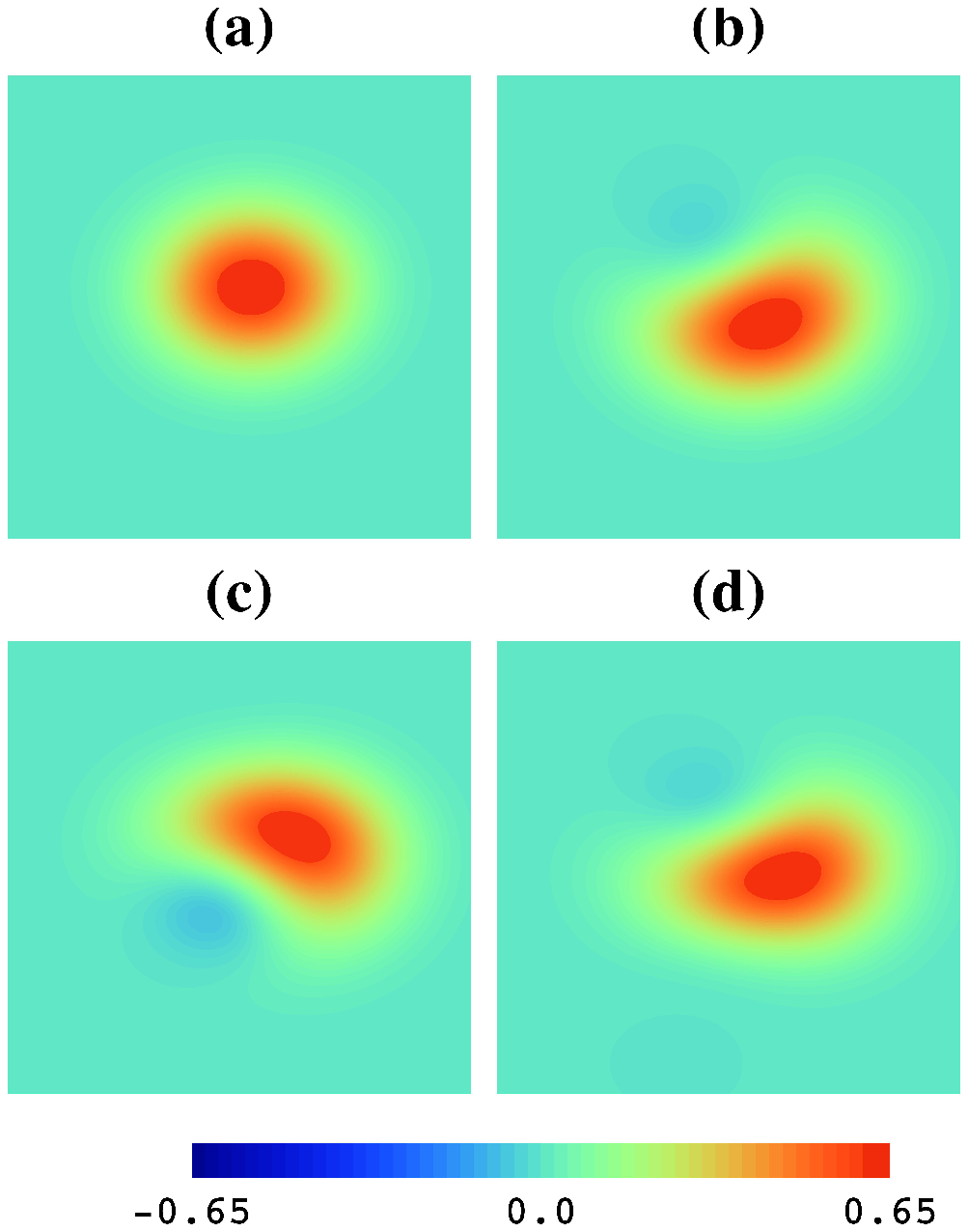}
\protect\caption{\label{fig:Wig-PSSC} (Color online) Wigner function of PSSCS for coherent state parameter $\alpha={\frac{1}{5}}\exp\left(\frac{i \pi}{3}\right)$ for photon addition of 0-3 in (a)-(d), respectively. Here, the squeezing parameters $r=0.1$ and $\phi=0$.}
\end{figure}

The Wigner function of PSSCS obtained in Eq. (\ref{eq:W-PSSC}) is shown in the contour plots of Figure \ref{fig:Wig-PSSC} with a repetitive photon subtraction from 0-3. Here, the perfect circular rings observed in Figure \ref{fig:Wig-PASC} are not present. The nonclassicality present in the state is captured in the negative values of Wigner function. 
\color{black}

\begin{figure}
\centering{}\includegraphics[scale=0.6]{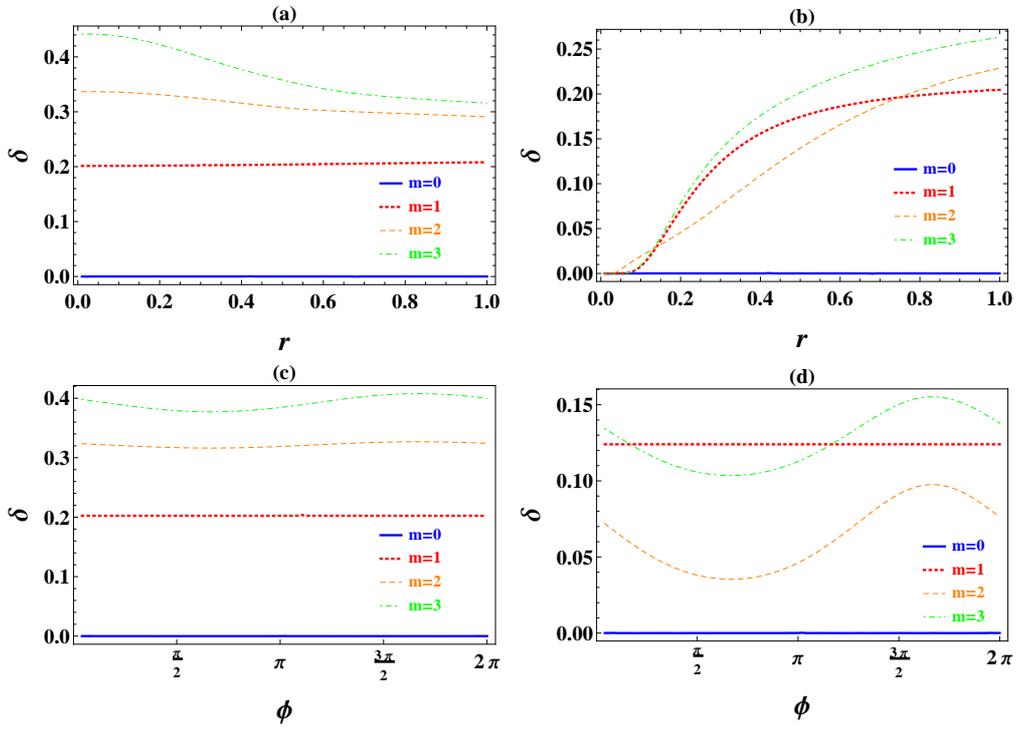}
\protect\caption{\label{fig:Wig-vol} (Color online) Wigner volume of PASCS (in (a) and (c)) and PSSCS (in (b) and (d)) is shown for coherent state parameter $\alpha={\frac{1}{5}}\exp\left(\frac{i \pi}{3}\right)$. In (a) and (b), variation is shown with the squeezing parameter $r$ for $\phi=0$. Dependence on the squeezing angle $\phi$ is shown in (c) and (d) considering $r=0.3$.}
\end{figure}

Until now, we have compared  the nonclassical character of PASCS and PSSCS using Wigner function and other witnesses of nonclassicality. Now, we would like  to provide a quantitative character to the investigation by computing a quantitative measure of nonclassicality which is known as nonclassical volume and defined as
 \cite{zyco} 
\begin{equation}
\delta=\frac{1}{2}\left(\int\int\left|W\left(\gamma,\gamma^{*}\right)\right|d^2\gamma-1\right).\label{eq:ncv}
\end{equation}
As nonclassical volume essentially measures the volume of the negative part of the Wigner function, a non-zero value of $\delta$ corresponds to nonclassicality, and the more is the value, the more nonclassical the state is. Variations of the amount of nonclassicality with noise and state parameters have been studied earlier in various physical systems using $\delta$ (cf. \cite{QDs,with-Jay} and references therein). Following the strategy adopted in the earlier works, in Figure \ref{fig:Wig-vol}, variation of $\delta$ obtained for PASCS and PSSCS are shown with squeezing parameters. From the figure, it may be concluded that nonclassicality increases with the number of photons added (cf. Figure \ref{fig:Wig-vol} (a)). However, no such conclusion regarding PSSCS can be made (cf. Figure \ref{fig:Wig-vol} (b)). Further, the nonclassicality obtained in both the states depends on the phase in the squeezing parameter.
\color{black}

\section{Conclusion}\label{sec:conclusion} 

We have calculated the nonclassical properties of photon added and photon subtracted squeezed coherent states. Employing the technique of IWOP, we have obtained the normalization constants, photon number distributions, various moments and phase space distributions for these states. Specifically, we have obtained both lower order and higher order antibunching, sub-Poissonnian photon statistics and the Hong-Mandel type squeezing. Further, we have also studied the Klyshko's criteria of nonclassicality and the parameter of efficiency of the approximate single photon source using obtained photon number distribution.

We have also computed the Wigner function for both PASCS and PSSCS. Finally, we have also made an attempt to quantify the nonclassicality present in the states using Agarwal's $A_{3}$ parameter and nonclassical volume. 

The comparative study led to a conclusion that \textit{PASCS is more nonclassical than PSSCS}. It is established using the witnesses of nonclassicality (like, Wigner function, Klyshko's criteria) and a quantitative measure (nonclassical volume) that photon addition results in a host of quantum phenomena.  

Further, addition of photons in the SCS enhanced the nonclassical behavior captured through various witnesses of nonclassicality and quantified through nonclassical volume, which is observed to increase with the number of photons added. In contrast, no such direct conclusion regarding the effect of photon subtraction on the amount of nonclassicality can be made as it depends on  various other parameters, such as squeezing parameters, odd/even number of photons subtracted, too.  

\noindent\textbf{Acknowledgment:} AP
thanks Department of Science and Technology (DST), India for the support
provided through the project number EMR/2015/000393.

\end{document}